\documentstyle[twoside,fleqn,espcrc2,epsf]{article}

\newcommand{\AmS}{{\protect\the\textfont2
  A\kern-.1667em\lower.5ex\hbox{M}\kern-.125emS}}

\newcommand{\bd}{\begin{displaymath}}
\newcommand{\ed}{\end{displaymath}}
\newcommand{\be}{\begin{equation}}
\newcommand{\ee}{\end{equation}}
\newcommand{\bea}{\begin{eqnarray}} 
\newcommand{\eea}{\end{eqnarray}}
\newcommand{\bt}{\begin{tabular}}
\newcommand{\et}{\end{tabular}\newline}

\newcommand{\pslasht}{\not{\hspace{-0.001cm}p}}

\newcommand{\Dlr}{\stackrel{\leftrightarrow}{D}}
\newcommand{\Dl}{\stackrel{\leftarrow}{D}}
\newcommand{\Dr}{\stackrel{\rightarrow}{D}}

\newcommand{\Dlrsl}{\not{\hspace{-0.1cm}{\Dlr}}}
\newcommand{\pslash}{\not{\hspace{-0.08cm}p}}

\newcommand{\cswv}{c_{sw}}
\newcommand{\dst}{\displaystyle}
\newcommand{\half}{\frac{1}{2}}
\newcommand{\fourth}{\frac{1}{4}}

\title{
\vspace{-3.65cm}
{\normalsize DESY 97--181}    \\[-0.2cm]
{\normalsize HUB--EP--97/61}  \\[-0.2cm]
{\normalsize September 1997}  \\
\vspace{2.25cm}                           
Perturbative Renormalization of Improved Lattice Operators\thanks{Talk
presented by S. Capitani at Lattice 97, Edinburgh.}}

\author{
S.~Capitani\address{DESY~--~Theory Group, Notkestrasse 85, D-22607 Hamburg, 
Germany}, 
M.~G\"ockeler\address{Universit\"at Regensburg, Institut f\"ur Theoretische 
Physik, D-93040 Regensburg, Germany}, 
R.~Horsley\address{Humboldt-Universit\"at, Institut f\"ur Physik, 
Invalidenstrasse 110, D-10115 Berlin, Germany}, 
H.~Perlt\address{Universit\"at Leipzig, Institut f\"ur Theoretische Physik,
Augustusplatz 10/11, D-04109 Leipzig, Germany}, 
P.~Rakow\address{DESY-IfH Zeuthen, Platanenallee 6, D-15738 Zeuthen, Germany}, 
G.~Schierholz$^{\rm a,\rm e}$ 
and A.~Schiller$^{\rm d}$
}

\begin{document}

\begin{abstract}
We derive bases of improved operators for all bilinear quark currents up to 
spin two (including the operators measuring the first moment of DIS 
Structure Functions), and compute their one-loop renormalization constants 
for arbitrary coefficients of the improvement terms. 
We have thus control over $O(a)$ corrections, and for a suitable choice of 
improvement coefficients we are only left with errors of $O(a^2)$.
\end{abstract}

\maketitle

\section{INTRODUCTION}

In this talk we extend previous calculations of the renormalization
constants of quark bilinear operators~\cite{mio,prev} using improved 
(Sheikholeslami-Wohlert~\cite{sw}) fermions. 
Though the calculations are performed in one-loop perturbation theory,
we quote results for arbitrary coefficients of the improved action and 
operators. If they are properly determined (for instance by imposing Ward 
identities~\cite{paulroger}), then only $O(a^2)$ errors are left.

\section{QUARK PROPAGATOR}

Let us first consider the quark propagator
\bd
S^{-1} = {\rm i} \pslash + m + a r p^2 /2 - \Sigma^{\rm latt},
\ed
where the bare mass is given by $m a = \frac{1}{2 \kappa} - 4 - 
\frac{g^2}{16\pi^2} C_F \Sigma_0$. We write the self-energy as
\vspace{-0.2cm}
\bea
\Sigma^{\rm latt} & = & \frac{g^2}{16\pi^2} C_F 
({\rm i}\pslash \Sigma_1^{\rm latt} + m \Sigma_2^{\rm latt} 
+ a r p^2 \Sigma_3^{\rm latt} \nonumber \\
& + & a r m {\rm i}\pslash \Sigma_4^{\rm latt} + a r m^2 \Sigma_5^{\rm latt} 
 + O(a^2)) . \nonumber
\eea 
In the covariant gauge we obtain (for $r=1$)\footnote{
We denote the coefficient of the improvement term in the action by $c_{sw}$. 
We use the abbreviations $\eta=1-\alpha$ (where $\alpha$ is the usual gauge 
parameter) and $L(ap) = \log(a^2p^2) + \gamma_E - F_0$. The quantities 
$\gamma_1 =  2$ and $\gamma_2 =  8 $ are anomalous dimensions, and 
$F_0 = 4.369225$.}:

\bea
\Sigma_0^{\rm latt} &=& -51.4347 + 13.7331 c_{sw} + 5.7151 c_{sw}^2 \nonumber\\
\Sigma_1^{\rm latt} &=&  15.6444 - 2.2489 c_{sw} - 1.3973 c_{sw}^2 \nonumber\\
&&+ \gamma_1 (1-\eta) \half L(ap)
+\eta \nonumber\\
\Sigma_2^{\rm latt} &=& 9.0680 - 9.9868 c_{sw}  - 0.0169 c_{sw}^2 \nonumber\\
&&+(\gamma_2 - \gamma_1 \eta ) \half L(ap) +2 \eta \nonumber\\ 
\Sigma_3^{\rm latt} &=& 7.0670 + 0.4857 c_{sw} - 0.0817 c_{sw}^2 \nonumber\\
& & + 0.0719 \eta + (-1 + 3 c_{sw} - 2 \eta ) \half L(ap)  \nonumber\\ 
\Sigma_4^{\rm latt} &=& 
 -6.2029 - 1.4850 c_{sw} + 1.2860 c_{sw}^2 \nonumber\\
& & - 0.1437 \eta + (-5 - 3 c_{sw} + 2 \eta ) \half L(ap) \nonumber\\ 
\Sigma_5^{\rm latt} &=& -13.4623 +16.9857 c_{sw} -1.5234  c_{sw}^2 \nonumber\\
& & - 2.0719 \eta +(-10 + 6 c_{sw} + \eta ) \half L(ap) \nonumber.
\eea
From this we derive the renormalization constant
\bd
Z_m = 1 -\frac{g^2}{16\pi^2}C_F\!(6 \log (a\mu) 
\ed
\bd
\quad\quad\quad - 12.952 - 7.738 c_{sw} + 1.380 c_{sw}^2) ,
\ed
and the critical $\kappa$
\be
\kappa_c(g)=\half(4 - \frac{g^2}{16\pi^2} C_F \Sigma_0)^{-1}.\label{eq:kappac}
\ee
In Eq.~(\ref{eq:kappac}) one has the choice of using $c_{sw}=1$ or the 
actual value used in true simulations. In Fig.~1 we compare the various 
choices with the data, and we find that the latter choice, combined with 
tadpole improvement, agrees best. This justifies our procedure. 

\begin{figure}[hbt]
\hspace*{1.0cm}
\epsfxsize=6.0cm \epsfbox{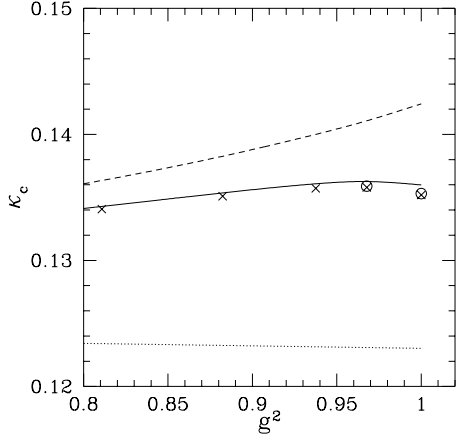}
\caption{The dotted and dashed curves correspond to $c_{sw}=1$, 
without and with tadpole improvement. The solid curve corresponds to $c_{sw}$ 
as given by the Alpha Collaboration~\cite{alpha}, plus tadpole improvement. 
This is compared with recent data.}
\end{figure}

The free propagator at $\kappa_c=\frac{1}{8r}$ is given by
$S(p) = \frac{1}{{\rm i}\pslasht} + \frac{a r }{2} + O(a^2)$.
The $O(a)$ term is constant in momentum space, and there are 
deviations from the continuum propagator only at very short distances. 
We thus suggest our Ansatz for the interacting propagator as:
\bd
 S(p) = \frac{Z_2}{1 + b_2 a r m} \frac{1}{{\rm i} \pslash + m_R} +a r \lambda.
\ed
Imposing this structure for the propagator, we find for the values of the
parameters (for $r=1$):
\bd
\lambda = \frac{1}{2} ( 1 + \frac{g^2}{16\pi^2} C_F 
( 2 \Sigma_1 - 2 \Sigma_3 ) )
\ed
\bd
Z_2 = ( 1 + \frac{g^2}{16\pi^2}  C_F \Sigma_1 ) 
\ed
\bd
b_2 = 1 + \frac{g^2}{16\pi^2} C_F ( 
  2 \Sigma_1 - \Sigma_2 - 2 \Sigma_3 - \Sigma_4 )
\ed
\bd
m_R = m Z_m ( 1 - b_m a r m )
\ed
\bd
\quad\quad = m ( 1 + \frac{g^2 }{16\pi^2} C_F ( \Sigma_1 - \Sigma_2 ) ) 
 ( 1 - \frac{a r m}{2} -
\ed
\bd
\quad  -\frac{a r m}{2} \frac{g^2}{16\pi^2} C_F ( 
 2 \Sigma_1 -\Sigma_2 - 2\Sigma_3 - 2\Sigma_4 + 2\Sigma_5 ) ) .
\ed
We see that $\lambda$ can be a constant only if $c_{sw} = 1$:
\bd
2 \Sigma_1 - 2 \Sigma_3 = 17.1548 - 5.4691 c_{sw} - 2.6311 c_{sw}^2
\ed
\bd
\quad\quad\quad + 1.8562 \eta + 6 (1 - c_{sw}) \half L(ap); 
\ed
for every other value of $c_{sw}$ there are $a \log(a p)$ 
contributions to $S(p)$.
The same happens for the parameters $b_2$ and $b_m$: 
both of them involve $a \log(a p)$ terms, that at $c_{sw} = 1$ 
cancel out. 

Two possible expressions that remove $O(a)$ effects from the quark propagator 
are:
\be
 S_{{\rm imp}}(p) \equiv (1 + b_2 a r m) ( S(p) - a r \lambda ) ;
\label{eq:one}
\ee
\be
S^{-1}(p) = (1 + b_2 a r m) S^{-1}_{{\rm imp}}(p) \label{eq:two}
\ee
\bd
\quad\quad\quad\quad\quad\quad 
- a r \lambda  S^{-1}_{{\rm imp}}(p) S^{-1}_{{\rm imp}}(p).
\ed
The improved propagators in these two definitions differ by terms 
of $O(a^2)$, but both are free of $O(a)$ effects.
Eq.~(\ref{eq:two}) is a non-linear equation which has to be solved 
iteratively, but it seems a better definition in practice than the first one.
This is because Eq.~(\ref{eq:one}) seems to have larger $O(a^2)$ effects,
thus using Eq.~(\ref{eq:two}) we can reach higher momenta.

\section{BASES FOR IMPROVED OPERATORS}

Some fundamental bases necessary to achieve full $O(a)$ 
improvement for point operators are:
\bd
\!\!(\bar{\psi}\psi)^{\rm imp} = (1+a \,b\, m)\bar{\psi}\psi - 
\half a c_1 \bar{\psi}\Dlrsl\psi
\ed
\bd
\!\!(\bar{\psi}\gamma_5\psi)^{\rm imp} = (1+a \,b\, m)\bar{\psi}\gamma_5\psi 
+ \half a c_1 \partial_\mu (\bar{\psi}\gamma_\mu\gamma_5\psi)
\ed
\bd
\!\!(\bar{\psi}\gamma_\mu\psi)^{\rm imp} = 
(1+a\, b\, m)\bar{\psi}\gamma_\mu\psi - 
\half a c_1 \bar{\psi}\Dlr_\mu \psi
\ed
\bd
\quad\quad\quad\quad+ {\rm i} \half a c_2 \partial_\lambda
(\bar{\psi}\sigma_{\mu\lambda}\psi)
\ed
\newpage
\bd
\!\!(\bar{\psi}\gamma_\mu\gamma_5\psi)^{\rm imp} =  
(1+a\, b\, m)\bar{\psi}\gamma_\mu\gamma_5\psi
\ed
\bd
\quad\quad\quad\quad- {\rm i} \half a c_1 \bar{\psi}\sigma_{\mu\lambda}
\gamma_5\Dlr_\lambda \psi + \half a c_2 \partial_\mu(\bar{\psi}\gamma_5\psi),
\ed
where $\Dlr_{\mu}=\Dr_{\mu}-\Dl_{\mu}$.
For the tensor operator see Ref.~\cite{nostro}. The normalizations are chosen
such that $c_i=1 + O(g^2)$ (in this way, for $g=0$ we realize tree-level 
improvement).

Some possible bases for the improvement of the one-link DIS operators
${\cal O}_{\mu\nu}=\bar{\psi}\gamma_\mu \Dlr_\nu \psi$ and the polarized
${\cal O}_{\mu\nu}^5=\bar{\psi}\gamma_\mu\gamma_5 \Dlr_\nu \psi$ are:
\be
{\cal O}_{\mu\nu}^{\rm imp} = 
( 1+ a\, b\, m)\bar{\psi}\gamma_\mu \Dlr_\nu \psi \label{eq:unpol}
\ee
\bd
\quad\quad - a\, c_1\, g \,
\bar{\psi}\sigma_{\mu\lambda} F_{\nu\lambda}^{\rm clover}\psi 
-\fourth\,a \,c_2 \, \bar{\psi}\{\Dlr_\mu,\Dlr_\nu\}\psi
\ed
\bd
\quad\quad  + \half\,a \,{\rm i}\,c_3 \partial_\lambda
(\bar{\psi}\sigma_{\mu\lambda}\Dlr_\nu\psi);
\ed
\be
{\cal O}_{\mu\nu}^{5,\rm imp} =
(1+ a\, b\, m)\bar{\psi}\gamma_\mu\gamma_5 \Dlr_\nu \psi \label{eq:pol}
\ee
\bd
\quad\quad  - a\, {\rm i} c_1\, g \, \bar{\psi}
\gamma_5 F_{\mu\nu}^{\rm clover}\psi 
\ed
\bd
\quad\quad -\fourth\,a \,{\rm i} \,c_2\,\bar{\psi}\sigma_{\mu\lambda}\gamma_5
\{\Dlr_\lambda,\Dlr_\nu\}\psi
\ed
\bd 
\quad\quad 
+ \half\,a \,{\rm i}\,c_3 \partial_\mu (\bar{\psi}\gamma_5\Dlr_\nu\psi). 
\ed
The relation $[\Dlr_\mu,\Dlr_\nu]^{\rm latt} = 
4 \,{\rm i} \,g \, F_{\mu\nu}^{\rm clover} + O(a^2)$ is  
useful to derive other bases.

If all these operators are inserted into forward matrix elements the
surface terms $\partial_\mu (\bar{\psi}{\cal O}\psi)$ vanish due to momentum 
conservation. Using the equations of motion it is also possible to further 
reduce the number of improvement coefficients.

\section{POINT OPERATORS}

We present the general $Z$ renormalization factors with the improvement
coefficients $c_{sw}$ (for the action) and $c_i$ (for the operators) kept 
general. These factors will be essential when a determination of the 
improvement coefficients is done~\cite{paulroger}.

Writing (to order $g^2$) $\langle{\cal O}\rangle |_{g^2}={\cal O}^{(1)} + a\, 
{\cal O}^{(2)}$,  the calculation of the amputated matrix elements for
$A_5 = \bar{\psi}\gamma_5 \psi$ and 
$A_\mu = \bar{\psi}\gamma_\mu\gamma_5 \psi$ gives

\bd
A_5^{(1)} =  \frac{g^2}{16\pi^2}C_F \gamma_5 (
0.5750 + 3.4333 c_{sw}^2
\ed
\bd
\quad\quad -2 \eta +(-4+\eta) L(ap) )
\ed
\bd
A_5^{(2)} =  0
\ed
\bd
A_\mu^{(1)} =\frac{g^2}{16\pi^2}C_F (\gamma_\mu\gamma_5 (
0.1519 -\eta - 19.3723 c_1
\ed
\bd
\quad\quad + 2.4967 c_{sw}+ 10.3167 c_1 c_{sw} -  0.8541 c_{sw}^2 
\ed
\bd
\quad\quad - 0.8846 c_1 c_{sw}^2 -(1-\eta)L(ap)) 
-2(1-\eta) \frac{\pslasht\gamma_5 p_\mu}{p^2})
\ed
\bd
A_\mu^{(2)} = \frac{{\rm i} g^2}{16\pi^2}C_F 
(\frac{1}{2}(\pslash\gamma_\mu\gamma_5 + \gamma_\mu\gamma_5\pslash)) 
\cdot(1.5323 
\ed
\bd
\quad\quad + 0.7322 c_1  - 1.7181 c_{sw} + 0.5430 c_1 c_{sw}
\ed
\bd
\quad\quad + 0.1302 c_{sw}^2 + 0.0537 c_1 c_{sw}^2 -\eta (0.8563
\ed
\bd
\quad\quad -4.0583 c_1) + (1-\eta)(1+ c_1)L(ap) ) .
\ed
Similar results hold for the other bilinears~\cite{nostro}. 
In general the $O(a)$ terms do not contribute to the tree-level structure 
in the massless case.

To extract the $Z$ factors we project forward matrix elements
onto their tree-level structure: \linebreak
$\langle q(p)|{\cal O}(\mu) |q(p) \rangle =
Z_{{\cal O}} Z_\psi^{-1} \; \langle q(p)|{\cal O}(a) |q(p) \rangle$,
and
$\langle q(p)|{\cal O}(\mu) |q(p) \rangle |_{p^2=\mu^2} =
\langle q(p)|{\cal O}(a) |q(p) \rangle |^{\rm tree}_{p^2=\mu^2}$,
where $Z_\psi$ is the wave function renormalization factor.
The one-loop renormalization factor $Z_{\cal O}$ for a lattice 
operator ${\cal O}$ can be cast into the form
\bd
Z_{\cal O}(a\mu,g) = 1-\frac{g^2}{16\pi^2} C_F (\gamma_{\cal O} 
\log (a\mu) +B_{\cal O}) ,
\ed
where $\gamma_{\cal O}$ is its anomalous dimension, and $B_{\cal O}$ 
the finite part of $Z_{\cal O}$.

The Wilson coefficients are usually computed in the $\overline{{\rm MS}}$ 
scheme; to convert the finite parts to this scheme one can use the relations
$B_{\cal O}^{MS} = B_{\cal O} - B_{\cal O}^{\rm cont}$ and 
$B_{\cal O}^{\overline{MS}} = B_{\cal O}^{MS} +
\frac{\gamma_{\cal O}}{2}(\gamma_E - \log (4\pi))$, 
where $\gamma_{\cal O}$ and $B_{\cal O}^{\rm cont}$, 
the finite contributions to  the continuum renormalization factors 
$Z_{\cal O}^{\rm cont}$, are:

\vspace{0.5cm}
\noindent \begin{tabular}{||c|c|c||}
\hline
${\cal O}$ &  
$\gamma_{\cal O}$ &  
$B_{\cal O}^{\rm cont}$ \\
\hline
$1$, $\gamma_5$  & 
$-6$ & 
$5 + \frac{\dst\gamma_{\cal O}}{\dst 2} 
(\gamma_E - \log (4\pi)) - \eta$ \\
[0.7ex]
$\gamma_\mu$, $\gamma_\mu\gamma_5$ & 
$0$ & 
$0$ \\
[0.7ex]
$\sigma_{\mu \nu} \gamma_5 $ & 
$ 2 $ & 
$-1 + \frac{\dst\gamma_{\cal O}}{\dst 2}
(\gamma_E - \log (4\pi)) + \eta$ \\
[0.7ex]
\hline
\end{tabular}

\vspace{0.5cm}
The results in the $\overline{{\rm MS}}$ scheme for all point operators
are then as follows:  
\newpage
\bd
B^{\overline{MS}}_1 = 12.9524 - 19.1718\;c_1 + 7.7379\;\cswv
\ed
\bd
\quad + 13.8007\;c_1\;\cswv - 1.3804\;\cswv^2 - 3.5383\;c_1\;\cswv^2
\ed
\bd
B^{\overline{MS}}_{\gamma_5}  =  22.5954 - 2.2487\; \cswv 
+ 2.0360\;\cswv^2
\ed
\bd
B^{\overline{MS}}_{\gamma_\mu}  =  20.6178  -9.7864\;c_1 - 4.7456\;\cswv
\ed
\bd
\quad + 3.4164\;c_1\;\cswv - 0.5432\;\cswv^2 + 0.8846\;c_1\;\cswv^2
\ed
\bd
B^{\overline{MS}}_{\gamma_\mu\gamma_5}  =  15.7963 - 19.3723\;c_1
+ 0.2478\;\cswv
\ed
\bd
\quad + 10.3167\;c_1\;\cswv - 2.2514\;\cswv^2 - 0.8846\;c_1\;\cswv^2
\ed
\bd
B^{\overline{MS}}_{\sigma_{\mu\nu}\gamma_5}  = 17.0181 - 16.2438\;c_1 
- 3.9133\;\cswv
\ed
\bd
\quad + 6.8553\;c_1\;\cswv - 1.9723\;\cswv^2 + 0.5897\;c_1\;\cswv^2.
\ed
With the appropriate values of $c_{sw}$ and $c_i$, all these 
operators are then fully $O(a)$ improved.

\section{ONE-LINK OPERATORS}

The many-link operators are essential in the OPE expansion of current-current 
correlators occurring in Structure Functions computations. Using
Eq.~(\ref{eq:unpol}) with $b=0$ and $c_3=0$ for the unpolarized case, 
in one-loop perturbation theory for the representation $\tau_1^{(3)}$ 
$({\cal O}_{\{44\}}-1/3 ({\cal O}_{\{11\}}+{\cal O}_{\{22\}}
+{\cal O}_{\{33\}}))$ we have:
\bd
\!\!{\cal O}_{\mu\nu,\tau_1^{(3)}}^{\rm imp} : B_{\cal O} = 
 -1.8826 - 3.9698 \cswv
\ed
\bd
\quad\quad\quad\quad - 1.0398  \cswv^2 +c_1 (5.9970 - 3.2685 \cswv)
\ed
\bd
\quad\quad\quad\quad - c_2 (6.6733 - 4.5371 \cswv - 0.4462 \cswv^2).
\ed

For the other representation, $\tau_3^{(6)}$ (${\cal O}_{\{14\}}$), 
we find full agreement with the $O(a)$ improved result given in 
Ref.~\cite{mio} for $c_{sw}=1$ and $c_1=c_2=1$.

For the polarized case, considering Eq.~(\ref{eq:pol}),
the result for the representation $\tau_4^{(6)}$ $({\cal O}^5_{\{14\}})$ is

\bd
\!\!{\cal O}_{\mu\nu,\tau_4^{(6)}}^{5,\rm imp}  : B_{\cal O}=
-4.0988  - 1.3593 \cswv - 1.8926 \cswv^2
\ed
\bd
\quad\quad\quad -c_2 (27.5719 -16.1193\cswv +0.7570 \cswv^2 ).
\ed

\end{document}